\documentclass[one column, 10pt]{IEEEtran}
 \usepackage{epsfig,datetime}
\usepackage{graphics}
\usepackage{amsmath}
\usepackage{amssymb}
\usepackage{graphicx}
\usepackage{cite}
\usepackage{epstopdf}
\usepackage{flushend} 
\newtheorem{theorem}{\bf Theorem}
\newtheorem{lemma}{\bf Lemma}

\newtheorem{definition}{\bf Definition}
\newtheorem{corollary}{\bf Corollary}

\DeclareMathAlphabet{\mathssf}{OT1}{cmss}{m}{sl}

\newcommand{\m}[1]{\mathbf{#1}^m}

\newcommand{\lon}[1]{\ln\left(#1\right)}    
\newcommand{\mk}[1]{\mathbf{#1}_1^k}
\newcommand{\mn}[1]{\mathbf{#1}_1^n}

\newcommand{\indi}[1]{{1\hspace{-2.3mm}{1}}_{\left\{#1\right\}}}

\newcommand{\peblk}{P_e^{blk}}

\newcommand{\expect}[1]{\mathbb{E}\left[#1\right]}

\newcommand{\bm}{\ensuremath{\mathrm{bit}\text{-}\mathrm{meters}}}

\begin{document}
\sloppy
\title{``Information-Friction'' and its implications on minimum energy required for communication\footnote{This paper was presented in part at the IEEE International Symposium on Information Theory (ISIT) 2013, Istanbul, Turkey.}}
\author{
  \IEEEauthorblockN{Pulkit Grover,}
  \IEEEauthorblockA{ECE, Carnegie Mellon University\\
    Email: pulkit@cmu.edu} 
    
}

\maketitle


\begin{abstract}
Just as there are frictional losses associated with moving masses on a surface, what if there were frictional losses associated with moving information on a substrate? Indeed, many modes of communication suffer from such frictional losses. We propose to model these losses as proportional to ``\bm,'' \textit{i.e.}, the product of mass of information (\textit{i.e., the number of bits}) and the distance of  information transport. We use this ``information-friction'' model to understand fundamental energy requirements on encoding and decoding in communication circuitry. First, for communication across a binary input AWGN channel, we arrive at fundamental limits on \bm\;(and thus energy consumption) for decoding implementations that have a predetermined input-independent length of messages. For encoding, we relax the fixed-length assumption and derive bounds for flexible-message-length implementations. Using these lower bounds we show that the \textit{total} (transmit + encoding + decoding) energy-per-bit must diverge to infinity as the target error probability is lowered to zero. Further, the closer the communication rate is maintained to the channel capacity (as the target error-probability is lowered to zero), the faster the required decoding energy diverges to infinity. 

\end{abstract}

\section{Introduction}

\begin{figure}[htbp] 
   \centering
   \includegraphics[width=3.2in]{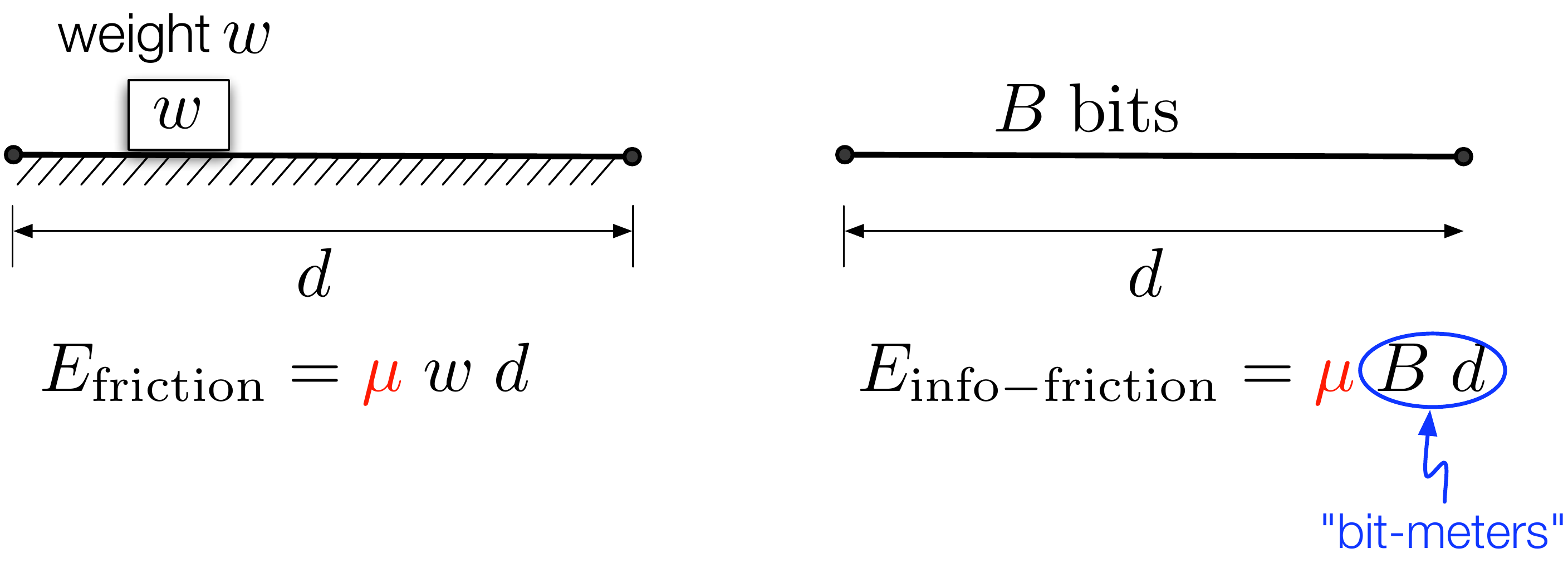} 
   \caption{A Newtonian inspiration for the information-friction model. The units of measuring energy are ``\bm,'' which is the product of number of bits of information, and the Euclidean distance to which that information travels, in the computation. }
   \label{fig:infofriction}
\end{figure}

Just as there are frictional losses associated with moving masses on a surface, there can be frictional losses associated with moving information between gates (see Fig.~\ref{fig:infofriction}) on a computational substrate. Within the context of communication, these frictional losses can be a significant part of the energy consumed in computations at the transmitter and the receiver (e.g.,~encoding and decoding an error-correcting code), which in turn can be a significant fraction of total energy for short-distance communication~\cite{JSAC11Paper}. 

What computational models allow us to account for these frictional losses? Communication complexity, introduced by Andrew Yao in~\cite{Yao}, accounts for information-movement on a computational substrate by counting the number of bits that need to be moved. However, for many implementations~\cite{ISIT12Paper} (as discussed in Section~\ref{sec:examples}), energy of computation depends not only on the number of bits, but also on the distance (Euclidean, \textit{i.e.}, $L_2$, or ``Manhattan''~\cite{JanBook}, \textit{i.e.}, $L_1$) to which those bits are moved. Are there models that account for these distances as well?

The VLSI model, introduced by Thompson and others in~\cite{thompson,thompsonthesis,BrentKung,chazelle,leiserson1981area,MeadConway} (and explored further in~\cite{sinhamultiply,kramerboolean,BhattUniversal,cole88,ThompsonSorting}), accounts for these distances by measuring the total wiring infrastructure required to compute a function. The product of the total wiring length and the number of clock-cycles needed, suitably scaled, is used as an approximation for energy consumed in computing. The required wiring infrastructure, as well as energy, are explored through upper and lower bounds (e.g.~\cite[Ch.~3 and Ch.~4]{thompsonthesis}). 

The focus on wires also limits the VLSI model in many ways. First, modern technology is exploring and using alternative interconnects (e.g.,~optical, carbon nanotubes, or even wireless~\cite{wirelessinterconnects}), and our nervous system uses axons and dendrites, none of which are made of metal wires, and can even evolve (if slowly) as the computation proceeds (e.g.~synapses in the brain and wireless interconnects)~\cite{DayanAbbott}. Second, modeling computational nodes as ones having small degree of connectivity, as is the case in the VLSI model~\cite{thompsonthesis}, can be too limiting. Third, even for metal-interconnects, the VLSI model focuses more on the wiring infrastructure needed to move information than on the amount and the distance of information actually moved in the computation. This can overestimate the energy requirements: for instance, not all wires need to be charged and discharged in each clock-cycle, but the model estimates energy consumption based on this assumption\footnote{Thompson does acknowledge this shortcoming in his thesis~\cite{thompsonthesis}.}. Finally, the lengths of messages passed on wires can be different in response to the input of computation, and thus energy-costs can be input dependent. This energy-difference is not accounted for in Thompson's model.



In Section~\ref{sec:implementation}, we introduce the ``information-friction''  model of computation and energy consumption (see Fig.~\ref{fig:infofriction}) that partially addresses these limitations of the VLSI-inspired models. Besides overcoming the limitations addressed above, the model is also appealing because of its conceptual simplicity and fewer assumptions in comparison with the VLSI model. The information-friction model accounts for the cost of computing by counting the ``\bm'': the product of the number of bits, and the distance to which these bits are moved (summed over all computation links). A similar ``\bm'' metric was used as a measure of ``transport capacity'' supported \textit{by} a communication network in the work of Gupta and Kumar~\cite{GuptaKumar}. Here, we are interested in the opposite question: how many \bm\;are \textit{needed to support} a computation?

When is ``\bm'' an appropriate metric for circuit communication energy? The issue is discussed in depth in Section~\ref{sec:examples}, where we argue that for many realistic models of computation (including computation on VLSI circuits), the energy consumption in links in the computational network is well approximated as (or is lowered bounded by) $\mu\times \mathrm{bm}$ where $\mu$ is a constant called the \textit{coefficient of information-friction}, and $\mathrm{bm}$ are the \bm\;required for the computation.  
Despite its intuitive appeal and applications, the metric has its shortcomings and limitations, which are also discussed in Section~\ref{sec:examples}.

In Section~\ref{sec:decoding}, we use the implementation model and an AWGN-based hard-decision channel model to derive the \bm\;cost for decoding an error-correcting code. Intellectually, our work builds on work of El Gamal, Greene, and Pang~\cite{ElGamalArea} that uses the VLSI model, to estimate complexity (but not energy) of encoding and decoding an error correcting code. This work also builds on our own work~\cite{ISIT12Paper} where we derive tradeoffs between wiring area and number of clock-cycles within Thompson's VLSI-model. In this paper, we show that the required \bm\;for decoding can be no smaller than $\Omega\left(\sqrt{\log\frac{1}{\peblk}/P_T}\right)$, where $\peblk$ is the block-error probability, and $P_T$ is the transmit power (for a binary-input AWGN channel where the receiver makes a hard decision on the channel output before decoding, see Section~\ref{sec:chmodel}). We show a similar result for encoding under a stronger model of circuit implementation: where the scheduling of messages along the communication links is not predetermined, but can adapt itself to the input of computation. Taking a step further, we also establish that if the communication rate $R$ is maintained close to the channel capacity $C(P_T)$ even as the (block) error-probability $\peblk$ is driven to zero, the required per-bit energy  goes up at least as fast as $\Omega\left(\sqrt{\frac{ \log n}{\log\frac{1}{2p_{ch}}}}\right)$. Here, $n$ is the blocklength of the code, and $p_{ch}$ is the cross-over probability of the Binary-Symmetric Channel (BSC) over which the signal is being communicated. As is well known, $n\gtrsim \Omega\left(\frac{\log\frac{1}{\peblk}}{K(C-R)^2}\right)$ for some constant $K$ (that depends on $p_{ch}$), and thus diverges to infinity faster as the rate and channel capacity are brought close to each other.

What are the implications of these results on \textit{total} (transmit + computation) energy consumption in communication? Under the information-friction model, optimizing over $P_T$, we show that the total (transmit + decoding) energy per bit is at least $\Omega\left(\sqrt[3]{\log\frac{1}{\peblk}}\right)$. This means that for any implementation that experiences information-frictional losses, the total energy per bit must diverge to infinity as the error probability is driven to zero. Further, operating with bounded transmit power (e.g., by operating close to the Shannon limit) appears\footnote{In absence of good upper bounds (that are a work in progress), we are left with comparing the lower bounds on energy consumed by the two strategies, which can only offer suggestions on which strategy is more energy-efficient.} to incur larger costs: the total energy per-bit is at least $\Omega\left(\sqrt{\log\frac{1}{\peblk}}\right)$.


%
%
Our results on information-frictional energy for encoding and decoding, and total energy for communication, attempt to begin to fill a void in our understanding of energy required for communication. In a paper that is little-known within the information-theory community~\cite{LandauerComm}, Landauer argues that one \textit{can} communicate with arbitrarily small energy, paralleling his results on zero-energy reversible computation~\cite{LandauerComputation}. In order to do so, however,  Landauer observes that one needs to  lower friction and noise in the communication medium to effectively zero\footnote{Of course, from an engineering viewpoint, it makes little sense to think about energy of computing assuming friction and noise are (or can be made) negligible. However, Landauer's main goal was not to provide practically relevant limits to energy of computing (as he himself acknowledges in~\cite{LandauerComm}), but instead to understand and resolve the paradox of Maxwell's demon~\cite{MaxwellsDemonFinal}. This fictional demon is able to lower the thermodynamic entropy of a system seemingly without expending any energy, a violation of the Second Law of Thermodynamics, which would mean (among other ``calamitous'' conclusions) that perpetual motion machines can exist. A fundamental limit on energy required for communication with arbitrarily small friction and noise would resolve the paradox (because measurement can be viewed as communication of information from the source to the measuring device). Landauer's contention in~\cite{LandauerComm} is that no such limit can exist and thus the paradox cannot be resolved by alluding to energy costs of communication. Instead it is losses in \textit{erasing} information that (according to Landauer) resolve the paradox. We refer the interested reader  to~\cite{LandauerDefended,ReplyToLandauerDefended,NortonShookUp,ShenkerGivesUpOnLandauer,ManojPaper1,ManojPaper2,Gawedzki} for contemporary work on energy of communication and computing within the context of theoretical physics, and discussions on whether Landauer's principle indeed resolves the paradox.}, which however requires lowering the speed of computing (asymptotically) to zero to keep the system in thermodynamic equilibrium. From this perspective, information-theoretic works of Golay~\cite{Golay} and Verd\'{u}~\cite{VerduCostCapacity} derive capacity per-unit energy for various communication media (\textit{i.e.}, channels) that do have friction and noise, but implicitly assume  that computation at the transmitter and receiver is frictionless and noiseless (and hence is free). In this paper, we take a step forward by allowing frictional losses in both communication and computation media and derive lower bounds on energy, whilst still ignoring noise in computation for simplicity.

\section{System model and notation}

\subsection{Channel model}
\label{sec:chmodel}
We consider a point-to-point communication link. An information sequence of $k$ fair coin flips $\mk{b}$ is encoded into $2^{nR}$ binary-alphabet codewords $\mn{X}$. The rate of the code is therefore $R=\frac{k}{n}$ bits/channel use, which is assumed to be fixed. The codeword $\mn{X}$ is modulated using BPSK modulation and sent through an Additive White Gaussian Noise (AWGN) channel of bandwidth $W$, with $W$ channel uses per second. The decoder estimates the input sequence $\mk{\widehat{b}}$ by first performing a hard-decision on the received channel symbols before using these (binary) hard-decisions $\mn{Y}$ to decode the input sequence. The overall channel $\mn{X}\rightarrow\mn{Y}$ is therefore a Binary Symmetric Channel (BSC) with raw bit-error probability $p_{ch}:=\mathbb{Q}\left(\sqrt{\frac{\zeta P_T}{\sigma_z^2}}\right)$, where $\mathbb{Q}(x)=\int_{x}^\infty \frac{1}{\sqrt{2\pi}}e^{-\frac{t^2}{2}}dt$, $\zeta$ is the path-loss associated with the channel, $P_T$ is the transmit power of the BPSK-modulated signal, and $\sigma_z^2$ is the variance of the Gaussian noise in the hard-decision estimation. The encoder-channel-decoder system operates at an average block-error probability $\peblk$ given by
$\peblk =  \Pr\left(\mk{\widehat{b}}\neq \mk{b}\right)$.
\begin{definition}[Channel Model ($\zeta,\sigma_z^2$)]
Channel Model ($\zeta,\sigma_z^2$) denotes (as described above) a BSC($p_{ch}$) channel that is a result of hard-decision at the receiver across an AWGN channel of average transmit power $P_T$, path loss $\zeta$ and noise variance $\sigma_z^2$.
\end{definition}

\subsection{Implementation, computation, and energy  models}\label{sec:implementation}
The computation is performed using a ``circuit'' on a ``substrate.'' This section formally defines these terms allowing for decoding analysis in Section~\ref{sec:decoding}. 
\begin{definition}[Substrate]
A Substrate is a square $\mathrm{Sq}(l)$ of side $l$ in $\mathbb{R}^2$ with vertices at $(0,0)$, $(0,l)$, $(l,0)$, and $(l,l)$.
\end{definition}
\begin{definition}[$\mathrm{Square Lattice} (\lambda)$]
A $\mathrm{Square Lattice} (\lambda)$ is the collection of points $(s\lambda,t\lambda)\in\mathbb{R}^2$ for all $s,t\in\mathbb{Z}$. 
\end{definition}

\begin{definition}[$\mathrm{Grid (\lambda)}$]
$\mathrm{Grid (\lambda)}$ is the intersection of $\mathrm{SquareLattice}(\lambda)$ with the substrate $\mathrm{Sq}(l)$, that is, it is the set of the lattice-points of the square lattice that lie in the substrate. 
\end{definition}
The parameter $\lambda$ determines how close computational nodes in the circuit can be brought to each other, and depends on the technology of implementation. For large circuits, $\lambda \ll l$.

\begin{definition}[Circuit, computational nodes]
The substrate $\mathrm{Sq}(l)$ together with a collection $\mathcal{S}\subset \mathrm{Grid}(\lambda)$ of points (called \textit{computational nodes}, or simply \textit{nodes}) inside $\mathrm{Sq}(l)$, is called a \textit{Circuit}, and is denoted by $\mathrm{Ckt}=(\mathrm{Sq}(l),\mathcal{S})$.
\end{definition}
For instance, $\mathrm{Sq}(10\lambda)$ along with the set $\mathcal{S}=\{(\lambda,\lambda),(5\lambda,4\lambda)\}$ constitutes a Circuit. 

Nodes can be \textit{input nodes}, \textit{output nodes}, or \textit{helper nodes}. Physically, the nodes help perform the computation by computing functions of received messages. Each node is accompanied with a finite storage memory. Input nodes store the input of computation (one bit each; at the beginning of computation), output nodes store the output (one bit each; at the end of computation), and helper nodes help perform the computation. 
\begin{definition}[Subcircuit]
A subcircuit $\mathrm{SubCkt}_1=(F_1,\mathcal{S}_1)$ of a circuit $\mathrm{Ckt}=(Sq(l),\mathcal{S})$ is constituted by an open and convex subset $\mathrm{F}_1$ of $\mathrm{Sq}(l)$ and by the subset of computational nodes $\mathcal{S}_1= F_1\cap \mathcal{S}$.
\end{definition}
That is, all the computational nodes within the sub-substrate $F_1$ must lie in the subcircuit $\mathrm{SubCkt}_1$.

\begin{definition}[Link]
A (unidirectional) link connects two nodes in that it allows for noiseless communication between nodes in one direction. The messages are binary-strings. Each message is a function of all the messages (and the possible inputs) received at the transmitting node until the start of the message-transmission.
\end{definition}
In a circuit with $n$ nodes, there are $n(n-1)$ unidirectional links, which can be used more than once during a computation.
\begin{definition}[Communication on a circuit]
Computational nodes use messages received thus far in computation, and stored memory values, to generate messages that can be communicated to other nodes over links. 
\end{definition}

We now introduce two models of computation: those with fixed and flexible-length messages. For both, the order of messages passed between computational nodes is pre-determined, but for a flexible-message-length computation, the length of a message can depend on the computation input. 
\begin{definition}[Fixed-message-length computation (on a circuit)]
The computation starts with the arrival of the input of computation at the input nodes. Each input node stores one bit of the input. The computation then proceeds with communication of messages of predetermined size, \textit{i.e.}, the messages' size does not depend on the input of computation. Each message is a function of the messages that the transmitting computational node has received thus far in the computation (including one bit of the input if the transmitting node is an input node). At the end of the computation, the output is available in the memories of the output nodes.
\end{definition}


\begin{definition}[Flexible-message-length computation (on a circuit)]
\label{def:flex}
The computation is said to be flexible-message-length computation if the number of bits in a message on a link in the computation can depend on the input of computation. Nevertheless, the minimum message-length is assumed to be at least one bit. 
\end{definition}


A computation may use some or all of the communication links in the circuit. Each link can be used as many times as needed, and at each use, the message can be of any chosen size with the associated costs as described in the following definitions. 
\begin{definition}[$\bm$\;cost of a link and of a circuit]\label{def:bm}
The $\bm$\;cost of a \textit{link} in a computation $\mathrm{Comp}$ on a circuit $\mathrm{Ckt}$ is the product of the total number of bits carried by the messages on the link and the Euclidean distance between the nodes at the ends of the link. The $\bm$\;for the entire circuit $\mathrm{Ckt}$ is the sum of $\bm$\;for all the links in $\mathrm{Comp}$. 
\end{definition}
Fixing the order of messages (but not necessarily the length), along with making the minimum message-size one bit, makes sure that there's no free-of-cost ``silence''~\cite{OrlitskySilence} that can be used for communicating messages between nodes. 
 Since each message on a link contains at least one bit, and the link is at least $\lambda$ in length, the message costs at least $\lambda\;\bm$. 

When a flexible-message-length computation is executed, the $\bm$\;expended can depend on the input of computation. In such cases, we will often be interested in $\textit{average}$ \bm\;for a link or a computation, where the average is taken over the possible input realizations (with a specified distribution). 


\begin{definition}[\bm\;for a link within a subcircuit]\label{def:bmcount}
For a link that connects two nodes within a subcircuit in a computation $\mathrm{Comp}$, the \bm\;for that link \textit{within the subcircuit} is the same as the \bm\;for the link in the original circuit. However, if only one of the nodes lies within the subcircuit, then \bm\;for this link within the subcircuit is the product of the number of bits of the message passed along this link and the length of link from the node inside the subcircuit to the boundary of the subcircuit. \end{definition}

\begin{definition}[\bm\;for a subcircuit]
The \bm\;for a subcircuit $\mathrm{SubCkt}_1=(F_1,\mathcal{S}_1)$ in computation $\mathrm{Comp}$ is the sum of \bm\;for all the links within the subcircuit (wholly or partially, as defined in Definition~\ref{def:bmcount}), and is denoted by $\bm(\mathrm{SubCkt}_1)$.
\end{definition}
The definition also holds for \bm\;for the entire circuit.

\begin{definition}[Coefficient of information-friction ($\mu$)]
The coefficient of information-friction, denoted by $\mu$, characterizes the energy required for computation in our model. This energy is given by $E=\mu\times \mathrm{bm}$, where $\mathrm{bm}$ is the number of \bm\;expended in executing the given computation on a circuit.
\end{definition}
\begin{definition}[Implementation Model ($\lambda,\mu$)]
Implementation Model ($\lambda,\mu$) denotes the implementation model as described in this section with $\lambda$ being the minimum distance between computational nodes, and $\mu$ being the coefficient of information-friction. 
\end{definition}
The same implementation model can be used to execute a fixed or flexible-message-length computation.

\section{Lower bounds on  \bm\;and information-friction energy of encoding and decoding}
\label{sec:decoding}
To obtain lower bounds on \bm\;for encoding and decoding, similar to analysis in~\cite{ElGamalArea,ISIT12Paper,ITW12Paper}, we need to cut the circuit under consideration into many disjoint subcircuits. The following definitions and lemmas set up the technical background needed for circuit-cutting and ensuing analysis.

\begin{definition}[Disjoint subcircuits]
Two subcircuits $\mathrm{SubCkt}_1=(F_1,\mathcal{S}_1)$ and $\mathrm{SubCkt}_2=(F_2,\mathcal{S}_2)$ of a circuit $\mathrm{Ckt}=(\mathrm{Sq}(l),\mathcal{S})$ are said to be \textit{disjoint subcircuits} if $F_1\cap F_2 = \phi$, the null set. Similarly, $\{\mathrm{SubCkt}_i\}_{i=1}^{N_{\mathrm{subckt}}}$ are said to be mutually disjoint subcircuits if $F_i\cap F_j=\phi$ for every $i,j\in \{1,2,\ldots,N_{\mathrm{subckt}}\},i\neq j$.
\end{definition}
It follows that any two disjoint subcircuits cannot share computational nodes or communication links that connect two nodes \textit{within} one of the subcircuits. In fact, two disjoint subcircuits do not share \bm\;of computation:

\begin{lemma}
\label{lem:subckt}
Let $\{\mathrm{SubCkt}_i\}_{i=1}^{N_{\mathrm{subckt}}}$, where  $\mathrm{SubCkt}_i=(F_i,\mathcal{S}_i)$, be a set of mutually disjoint subcircuits of the circuit $\mathrm{Ckt}=(\mathrm{Sq}(l),\mathcal{S})$. Then for any computation $\mathrm{Comp}$,
\begin{equation}\label{eq:subckt}
\bm (\mathrm{Ckt}) \geq \sum_{i=1}^{N_\mathrm{subckt}} \bm (\mathrm{SubCkt}_i).
\end{equation}
\end{lemma}
\begin{IEEEproof}
The lemma follows from the observation that in Definition~\ref{def:bm}, no \bm\;are double-counted in disjoint subcircuits. We note that there are potential situations when $\bigcup_{i=1}^{N_\mathrm{subckt}}F_i = \mathrm{Sq}(l)$ for which~\eqref{eq:subckt} is not satisfied with equality. This happens when there is a long link in a circuit which has a part that does not lie within either of the subcircuits that contain the two nodes at the ends of the link. 
\end{IEEEproof}

\begin{figure}[htbp] 
   \centering
   \includegraphics[width=3.5in]{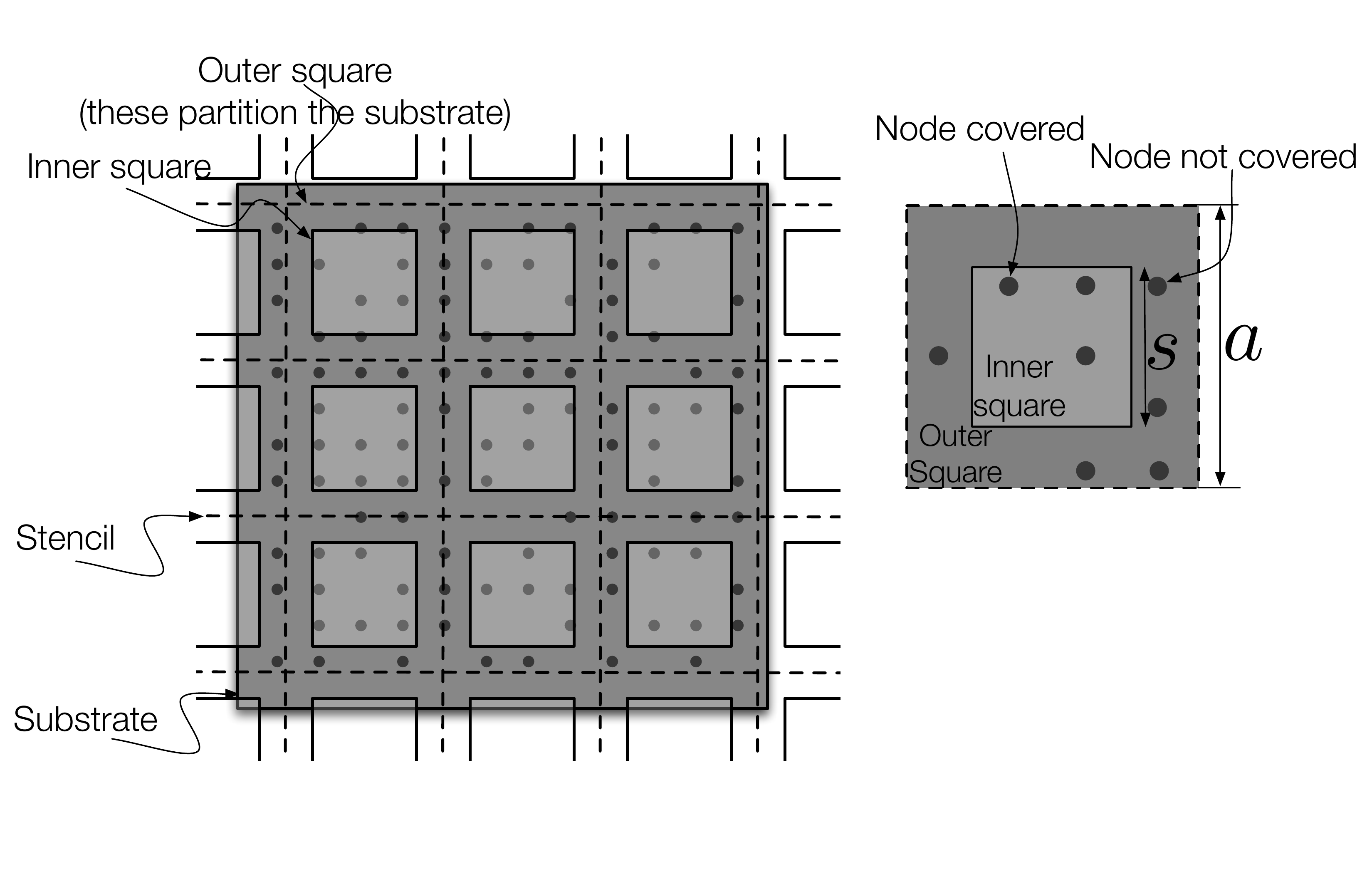} 
   
   \vspace{-0.3in}
   
   \caption{A Stencil overlaid on the Substrate. Also shown are the computational nodes of the Circuit on the Substrate. A zoomed-in version shows the dimensions of the Stencil. As an example, for the square in the zoomed-in version, $k_i^{\mathrm{inside}}=3$.}
   \label{fig:stencil}
\end{figure}

The decoder circuit is partitioned into multiple subcircuits via a ``Stencil\footnote{We use the term ``Stencil'' in analogy with the classic stencil instrument used to produce letters or designs on an underlying surface. A stencil can be slid on the surface to produce the design at any location on the surface, effectively shifting the origin-point of the design. In this case, a pattern of inner and outer squares is produced on the computational substrate.  
}'' that can be ``moved'' over the circuit by changing its origin.
\begin{definition}[Stencil]
A \textit{Stencil}$(a,\eta,O)$ in $\mathbb{R}^2$, for $\eta<\frac{1}{2}$, is a pattern of equally spaced ``inner'' squares that are concentric with ``outer'' squares which form a grid (as shown in Fig.~\ref{fig:stencil}). The length of a side of each outer square is $a$, and the origin $O$ lies in the center of an ``inner'' square. The side of each inner square is of length $s=(1-2\eta)a$.
\end{definition}

A node in a circuit is said to be \textit{covered} by a Stencil that is overlaid on the circuit substrate if it lies inside an inner-square of the Stencil. For the decoder, the $n$ input nodes store the channel observations, and the $k$ output nodes, also called ``bit-nodes,'' store the decoded message bits. At the encoder, the $k$ information-bits that are the input of computation are assumed to be stored in bit-nodes. Inside the $i$-th subcircuit, let $k_i^{\mathrm{inside}}$ denote the number of bit-nodes that lie inside the \textit{inner} square, and $n_i$ denote the number of input nodes that lie inside the \textit{outer} square (\textit{i.e.}, anywhere inside the $i$-th subcircuit).
\begin{definition}[Stencil-partition]
\label{def:partition}
The outer squares of \textit{Stencil}$(a,\eta,O)$ induce a partition (see Fig.~\ref{fig:stencil}) of a circuit into subcircuits, each occupying substrate area at most $a^2$. If any computational node lies on the boundary of an outer square, then it is arbitrarily included in one of the subcircuits.
\end{definition}

The next lemma shows that by moving the Stencil over the substrate, we can find at least one position of the Stencil so that the average number of nodes (over random locations of the Stencil) are covered.

\begin{lemma}
\label{lem:eta}
For any circuit implemented in Implementation Model ($\lambda,\mu$), for any $\eta>0$, there exists an origin $O$ of \textit{Stencil}$(a,\eta,O)$  such that the number of bit-nodes covered by the Stencil is lower bounded by
\begin{equation}
\sum_i k_i^{\mathrm{inside}}\geq k(1-2\eta)^2.
\end{equation}
\end{lemma}
\begin{IEEEproof}
The proof uses the probabilistic method~\cite{alonspencer}.  Let $O\sim \mathbb{U}\{[0,a),[0,a)\}$, that is, uniformly distributed in the square formed by $(0,0),(0,a),(a,a),(a,0)$. Now, the average number of bit-nodes covered by the Stencil (averaged over $O$) is:
\begin{eqnarray}
\nonumber\expect{\sum_{i=1}^k \indi{i\;covered}}&=&\sum_{i=1}^k \expect{\indi{i\;covered}}\\
\nonumber&=&\sum_{i=1}^k \Pr(i\;covered)\\
&\overset{(a)}=&\sum_{i=1}^k (1-2\eta)^2 \\
&=& k(1-2\eta)^2
\end{eqnarray}
where the key step $(a)$ follows from the observation that for any point, as we move the origin $O$ around uniformly, the probability measure of the set of origins for which the point is covered by the Stencil is the fraction of area covered by the Stencil, which is $(1-2\eta)^2$. Thus there exists at least one value of the origin $O$ such that the number of nodes covered is no smaller than the average.
\end{IEEEproof}

Consider the Stencil shown in Fig.~\ref{fig:stencil}. The distance between the inner and the outer squares is $\eta a$. $B$ bits are said to be \textit{communicated} from the ``transmitting'' part of the circuit to the ``receiving'' part if the values stored in the receiving part are independent of the $B$ bits prior to communication, and the bits can be recovered (in an error-free manner) from the messages \textit{received at the receiving} part during the process of communication. Notice that this definition is looser than the traditional understanding of communication: we do not stipulate that the \textit{stored values} at the receiving part post-communication be able to recover the $B$ bits. 

 If $B$ bits are communicated from outside an outer square to inside an inner square in a subcircuit, then, intuitively, the \bm\;associated with the subcircuit should be at least $\eta aB$. The following lemma shows this rigorously:
\begin{lemma}[\bm\;and average \bm\;in computations]
\label{lem:fixedschedule}
Consider a circuit implemented in Implementation Model ($\lambda,\mu$), and any subcircuit $\mathrm{SubCkt}$ obtained using the Stencil-partition defined in Definition~\ref{def:partition}. For communicating $B$ bits of information from outside an outer-square to inside the corresponding inner-square, $\bm\geq \eta aB$ for fixed-length messages. Further, even allowing for a flexible-message-length, the \textit{average} $\bm\geq \eta aB$. Similarly, for communicating $B$ bits from inside an inner-square to outside the corresponding outer-square, the average $\bm\geq \eta aB$.

\end{lemma}

\begin{IEEEproof}
\begin{figure}[htbp] 
   \centering
\includegraphics[width=3in]{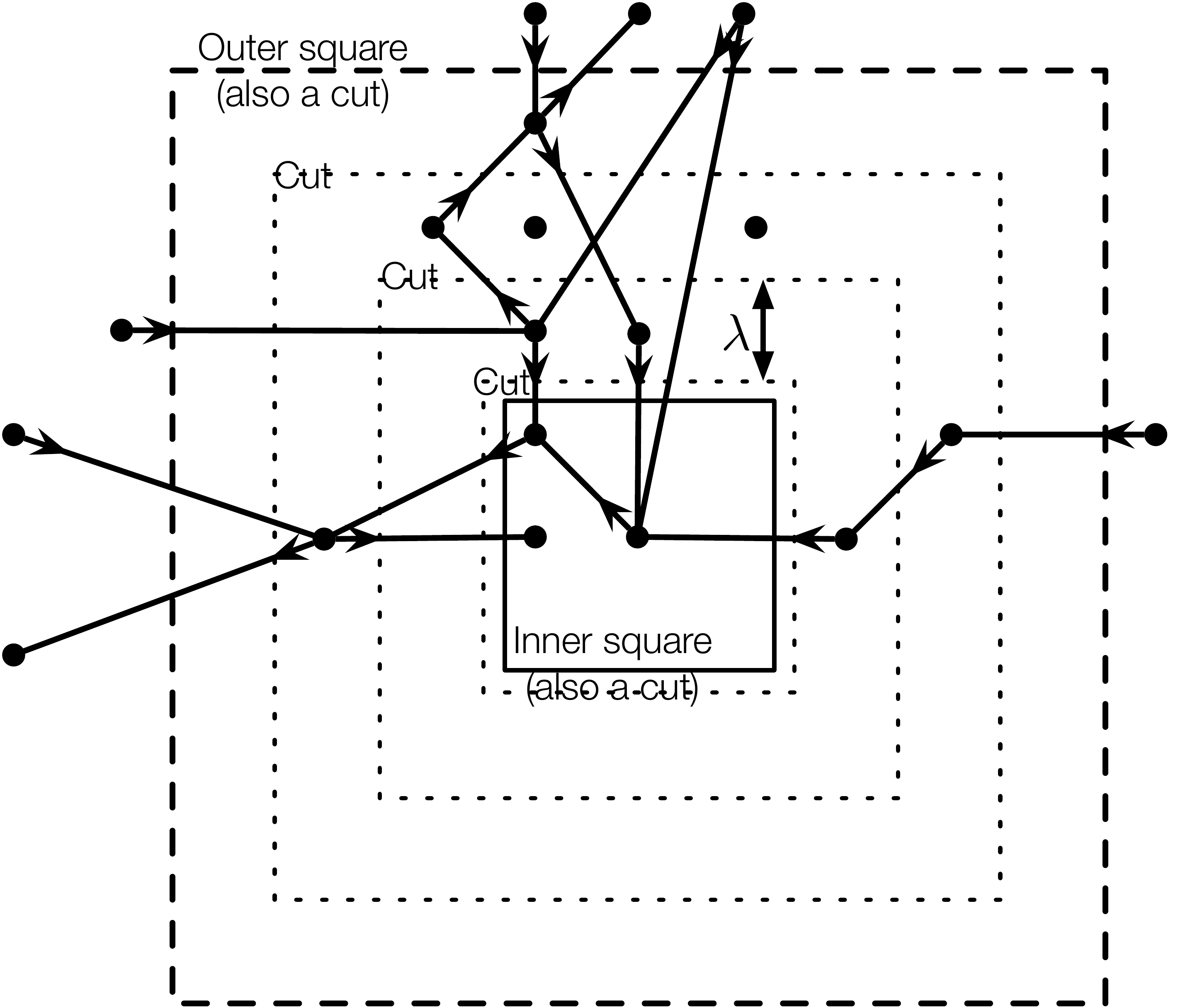}  
   \caption{Square cuts are made in order to use the cut-set bounding technique. The directed edges show the links along which information flows in the computation. However, the links do not indicate the relative order of information flow during the computation, or the amount of information they carry.}
   \label{fig:cuts}
\end{figure}

\textit{Fixed-length messages:} 
Consider the concentric $N_{cut}$ square-shaped cuts on the sub-circuit-network, starting with the outer square as a cut, with distance $\lambda$ separating these cuts, as shown in Fig.~\ref{fig:cuts}. The cuts end when distance from the inner square is smaller than $\lambda$. This remaining distance is denoted by $\alpha\lambda$ for some $\alpha\in [0, 1)$. The inner square is now included as the final $N_{cut}$-th cut. Except for the inner square, across each cut, each link has to cross at least $\lambda$ distance. 

Further, if the number of bits across any cut, which is the summation of bits passed over all links across the cut, is smaller than $B$, then $B$ bits cannot be delivered to the inner square. Thus across each cut, the total number of $\bm$ should be at least $B\lambda$. If $N_{cut}$ is the number of cuts, the total distance for which at least $B$ bits need to travel is at least $(N_{cut}-2)\lambda + \alpha\lambda$ which is exactly the distance $\eta a$ between the inner square and the outer square. Thus, for fixed-message-length computation, $\bm (\mathrm{SubCkt})\geq \eta a B$.

\textit{Flexible-message-length}: Flexible-message-length allows for use of variable-length messages on circuit links that can depend on the input of computation. Nevertheless, to code $B$ bits of information using variable-length coding still requires\footnote{We remind the reader that ``silence'' can not be used for communication because each message has at least one bit (see Definition~\ref{def:flex}).} at least $B$ bits \textit{on average}~\cite[Pg.~110]{CoverThomas}. 
\end{IEEEproof}
\subsection{Decoding lower bounds: fixed-length messages}

\begin{lemma}\label{lem:fanoex}
If at most $\frac{r}{3}$ bits of information is available to obtain an estimate $\widehat{M}$ of a variable $M$ that is distributed uniformly on the set $\mathcal{M}:=\{1,2,\ldots,2^{r}\}$, $r$ being a positive integer, then $\Pr(\widehat{M}\neq M)\geq\frac{1}{9}$.
\end{lemma}
\begin{IEEEproof}
Applying Fano's inequality~\cite[Pg.~39]{CoverThomas} to reconstruction of message $M$,  given the available information $I$ of at most $r/3$ bits, the error probability $P_e:=\Pr(\widehat{M}\neq M)$ is lower bounded by
\begin{eqnarray}
\nonumber &&P_e\log(|\mathcal{M}|-1)+h_b(P_e)\geq H(M|I) \\
\nonumber &=& H(M) - H(I) + H(I|M)\\
 &\geq & H(M)-H(I) \geq r-r/3 = 2r/3\label{eq:fano},
\end{eqnarray}
where $h_b(\cdot{})$ on the LHS is the binary entropy function. We now consider two cases:

\textit{Case 1:  $r= 1$}: In this case, $|\mathcal{M}|=2$ and $\log(|\mathcal{M}|-1)=0$, and thus from~\eqref{eq:fano},
\begin{equation}
\label{eq:case1}
h_b(P_e)\geq \frac{2}{3}.
\end{equation}
Since $h_b(x)\leq 2\sqrt{x(1-x)} \leq 2\sqrt{x}$ for $x\in(0,0.5)$ (see, e.g.~\cite{ITBlog}), $x\geq\frac{(h_b(x))^2}{4}$. From~\eqref{eq:case1}, for $r=1$,
\begin{equation}
P_e\geq \frac{(h_b(P_e))^2}{4}\geq \frac{4}{9\times 4}=\frac{1}{9}.
\end{equation}
\textit{Case 2:  $r \geq 2$}: In this case, $|\mathcal{M}|\geq 4$, and thus using a looser form of~\eqref{eq:fano},
\begin{eqnarray*}
P_e\log(|\mathcal{M}|)+h_b(P_e) \geq \frac{2r}{3}\\
\Rightarrow P_e\log(|\mathcal{M}|)+1 \geq \frac{2r}{3}\\
\Rightarrow P_e \geq \frac{\frac{2r}{3}-1}{\log(|\mathcal{M}|)}= \frac{\frac{2r}{3}-1}{r} = \frac{2}{3} - \frac{1}{r}\\
\overset{(r\geq 2)}\geq \frac{2}{3}-\frac{1}{2}=\frac{1}{6}>\frac{1}{9}.
\end{eqnarray*}
\end{IEEEproof}
We can now connect information-flow in decoding subcircuits to error probability. The following lemma provides a lower bound on the error probability when the number of \bm\;in a subcircuit of the decoder implementation is sufficiently small. 
\begin{lemma}\label{lem:bm}
For any decoder subcircuit $\mathrm{SubCkt}_i$ obtained via Stencil-partitioning of Implementation Model $(\lambda,\mu)$, with $k_i^{\mathrm{inside}}\geq 1$, if $\bm(\mathrm{SubCkt}_i)<\eta a\frac{k_i^{\mathrm{inside}}}{3}$, then $\peblk\geq \frac{(2p_{ch})^{n_i}}{9}$. 
\end{lemma}
\begin{IEEEproof}
From Lemma~\ref{lem:fixedschedule}, since the number of \bm\;for the subcircuit is smaller than $\eta a\frac{k_i^{\mathrm{inside}}}{3}$, and the distance between the outer square and the inner square is $\eta a $ meters, at most $\frac{k_i^{\mathrm{inside}}}{3}$ bits of information $I$ can be communicated from outside the outer square to inside the inner square. 

We first observe that a BSC($p_{ch}$) is a stochastically degraded version of a BEC($2p_{ch}$). That is, a decoder that receives channel outputs that pass through BEC($2p_{ch}$) can simulate a BSC($p_{ch}$) channel by randomly assigning the value $0$ or $1$ to an erased bit, \textit{i.e.} without any increase in \bm. Supplying the decoder with outputs of the erasure channel, we examine the event $\mathcal{E}$ when \textit{all} the $n_i$ channel outputs inside the outer square are erased. This event has  probability $(2p_{ch})^{n_i}$. 

Conditioning on the erasure event $\mathcal{E}$, let the (block) probability of not recovering all of the bits inside the $i$-th inner square, denoted by $\vec{b}_i^\mathrm{in}$, be $P_{e,i}^{\mathcal{E}}$. From Fano's inequality~\cite[Pg.~39]{CoverThomas} applied to reconstructing the message bits $\vec{b}_i^\mathrm{in}\in\mathcal{B}_i$, $|\mathcal{B}_i|=2^{k_i^{\mathrm{inside}}}$,  given the communicated information $I$ of entropy at most $k_i^{\mathrm{inside}}/3$ bits,
\begin{eqnarray}
P_{e,i}^{\mathcal{E}}>\frac{1}{9}.
\end{eqnarray}
Thus, for any $k_i\geq 1$, the (unconditional) error probability for recovering the $k_i^{\mathrm{inside}}$ bits correctly is lower bounded by $\frac{(2p_{ch})^{n_i}}{9}$. Since the block-error probability $\peblk$ for the entire code is larger than the block-error probability in recovering the $k_i^{\mathrm{inside}}$ bits in $i$-th subcircuit, we obtain the lemma.
%
\end{IEEEproof}


\begin{lemma}
\label{lem:choosinga}
For the Implementation Model  ($\lambda,\mu$), for Stencil-partition with outer-squares of side-length $a$, the maximum number of computational nodes (input, output, or helper) in a subcircuit is upper bounded by
\begin{equation}
\label{eq:choosinga}
N_{nodes}\leq \frac{a^2}{\lambda^2} + 4 \frac{a}{\lambda} + 4.
\end{equation}
Further, if $\frac{a^2}{\lambda^2}\geq 25$,
\begin{equation}
N_{nodes}\leq 2\frac{a^2}{\lambda^2}
\end{equation}
\end{lemma}
\begin{IEEEproof}
The number of nodes in a Stencil cell is approximately $\frac{a^2}{\lambda^2}$. The actual number could however be larger because of boundary effects. On each axis, allowing for one extra node to be included from either side of the square, the number of nodes is (loosely) upper bounded by $(\frac{a}{\lambda}+2)^2=\frac{a^2}{\lambda^2} + 4 \frac{a}{\lambda} + 4$. Also note that
\begin{eqnarray*}
\frac{2a^2}{\lambda^2}-(\frac{a}{\lambda}+2)^2 =\frac{a^2}{\lambda^2} - 4 \frac{a}{\lambda} - 4 = (\frac{a}{\lambda}-2)^2 - 8,
\end{eqnarray*}
which is positive (in fact, greater than $1$) when $\frac{a}{\lambda}\geq 5$, or $\frac{a^2}{\lambda^2}\geq 25$. 
\end{IEEEproof}

\begin{theorem}
\label{thm:decoding}
For an error correcting code transmitted over a channel with Channel Model ($\zeta,\sigma_z^2$) and decoded in a decoder circuit $\mathrm{DecCkt}$ implemented in Implementation Model ($\lambda,\mu$) with fixed-message-length implementation that achieves a block-error probability $\peblk$, the decoder \bm\;are lower bounded as:
\begin{equation}
\bm(\mathrm{DecCkt})\geq \frac{k}{48\sqrt{2}}\sqrt{\frac{\log\frac{1}{10 \peblk}}{\log\frac{1}{2p_{ch}}}}\lambda,
\end{equation}
\begin{equation}\label{eq:condition}
\text{as long as $\log\frac{1}{10 \peblk}>50\log\frac{1}{2p_{ch}}$.}
\end{equation}
\end{theorem}
\textit{Remark:} When condition~\eqref{eq:condition} is violated in the asmyptopia of $\peblk\to 0$, \textit{i.e.}, when 
\begin{equation}\label{eq:ineqa}50\log\frac{1}{2p_{ch}}\geq\log\frac{1}{10 \peblk},\end{equation} 
the transmit power $P_T$ needs to scale at least as fast as $\Omega\left(\log\frac{1}{\peblk}\right)$. To see this, we use a known bound~\cite{TateQFunction} on the $\mathbb{Q}$-function, namely, $\mathbb{Q}(x)\geq \frac{x}{1+x^2} \frac{e^{-x^2/2}}{\sqrt{2\pi}}$:
\begin{equation}
p_{ch}=\mathbb{Q}\left(\sqrt{\frac{\zeta P_T}{\sigma_z^2}}\right)\geq  \frac{ \sqrt{\frac{\zeta P_T}{\sigma_z^2}}}{1+ \frac{\zeta P_T}{\sigma_z^2}} \frac{e^{- \frac{\zeta P_T}{2\sigma_z^2}}}{\sqrt{2\pi}}.
\end{equation}
Thus,
\begin{eqnarray}
\lon{\frac{1}{p_{ch}}} &\leq &\lon{\sqrt{2\pi}\frac{1+\frac{\zeta P_T}{\sigma_z^2}}{\sqrt{\frac{\zeta P_T}{\sigma_z^2}}}}  + \frac{\zeta P_T}{2\sigma_z^2}\\
\label{eq:conditiona}&\overset{(a)}< & 2\frac{\zeta P_T}{\sigma_z^2}, \;\;\text{if $\frac{\zeta P_T}{\sigma_z^2}\overset{(b)}\geq 2$},
\end{eqnarray}
where $(a)$ follows from the observation that $\lon{\sqrt{2\pi}} + \lon{\frac{1+x}{\sqrt{x}}} + \frac{x}{2}< 2x$ for $x\geq 2$ (a fact that can be verified by simply plotting the two sides of the inequality). Further, if condition $(b)$ is not satisfied, then $P_T$ is bounded, and so is $p_{ch}$, which means that~\eqref{eq:condition} is not violated in the limit $\peblk\to 0$. From $(a)$ above and~\eqref{eq:ineqa}, $P_T=\Omega\left(\log\frac{1}{\peblk}\right)$ under condition $(b)$. This lower bound, which is derived for the case when condition~\eqref{eq:condition} is not satisfied, is larger than our lower bounds on total power when condition~\eqref{eq:condition} \textit{is} satisfied (Section~\ref{sec:total}). 


\begin{IEEEproof}
The outer squares of the Stencil partition the circuit into subcircuits. Let the $i$-th subcircuit have $n_i$ channel output nodes available within the \textit{outer} square and $k_i^{\mathrm{inside}}$ bit-nodes inside the \textit{inner} square.  Using Lemma~\ref{lem:eta}, we choose the origin $O$ of the Stencil so that at least $(1-2\eta)^2$ fraction of the $k$ bit-nodes are covered by the \textit{inner} squares, \textit{i.e.}, 
\begin{equation}
\sum_i k_i^{\mathrm{inside}}\geq (1-2\eta)^2k. 
\end{equation}
From Lemma~\ref{lem:choosinga} choosing Stencil parameter $a$ to be $\frac{1}{\sqrt{2}}\sqrt{\frac{\log\frac{1}{10\peblk}}{\log\frac{1}{2p_{ch}}}}\lambda$, under condition~\eqref{eq:condition},
\begin{eqnarray*}
\frac{a^2}{\lambda^2}= \frac{1}{2}\frac{\log\frac{1}{10\peblk}}{\log\frac{1}{2p_{ch}}}\overset{(\text{under}~\eqref{eq:condition})}>\frac{50}{2}=25. 
\end{eqnarray*}
Thus $\frac{a^2}{\lambda^2}\geq 25$. Using Lemma~\ref{lem:choosinga}, $n_i\leq \frac{2a^2}{\lambda^2} = \frac{\log\frac{1}{10\peblk}}{\log\frac{1}{2p_{ch}}}$.


From Lemma~\ref{lem:bm}, if \bm\;for any subcircuit are smaller than $\frac{k_i^{\mathrm{inside}}}3\eta a$, then the error probability is lower bounded as
\begin{eqnarray}
\peblk &\geq & \frac{(2p_{ch})^{n_i}}{9}\geq\frac{1}{9}(2p_{ch})^\frac{\log\frac{1}{10\peblk}}{\log\frac{1}{2p_{ch}}}=\frac{10}{9} \peblk,
\label{eq:contra}
\end{eqnarray}
which is a contradiction. 
%
Thus, for each decoding subcircuit $\mathrm{SubCkt}_i$ obtained via the Stencil-partition, 
\begin{eqnarray*}
\bm(\mathrm{SubCkt}_i)&\geq& \frac{k_i^{\mathrm{inside}}\eta a}{3}.
\end{eqnarray*}
From Lemma~\ref{lem:eta}, $\sum_{i}k_i^{\mathrm{inside}}\geq (1-2\eta)^2k$, therefore, using Lemma~\ref{lem:subckt},
\begin{eqnarray*}
&&\sum_{i=1}^{N_{\mathrm{subckt}}}\bm(\mathrm{SubCkt}_i) \geq  \frac{(1-2\eta)^2k\eta a}{3}\\
&=&\frac{(1-2\eta)^2k\eta }{3\sqrt{2}} \sqrt{\frac{\log\frac{1}{10\peblk}}{\log\frac{1}{2p_{ch}}}}\lambda.
\end{eqnarray*}
Choosing $\eta=\frac{1}{4}$ yields the theorem.
\end{IEEEproof}

\subsection{Encoding lower bounds: fixed and flexible-message-length}

\begin{theorem}
\label{thm:encoding}
For an error correcting code encoded in a circuit $\mathrm{EncCkt}$ that is implemented in Implementation Model ($\lambda,\mu$) and transmitted over a channel with Channel Model ($\zeta,\sigma_z^2$) and with block-error probability $\peblk$, the encoder average \bm\;(denoted by $\overline\bm$) are lower bounded as:
\begin{equation}
\overline\bm(\mathrm{EncCkt})\geq \frac{k}{48\sqrt{2}}\sqrt{\frac{\log\frac{1}{10 \peblk}}{\log\frac{1}{2p_{ch}}}}\lambda,
\end{equation}
\begin{equation}\label{eq:condition2}
\text{as long as $\log\frac{1}{10 \peblk}>50\log\frac{1}{2p_{ch}}$,}
\end{equation}
for both fixed and flexible-message-length encoding.
\end{theorem}

\begin{IEEEproof}
We directly show the result for flexible-message-length implementations, which subsume fixed-message-length implementations. At the encoder, $k$ input information bits are mapped to $n$ codeword output bits. 

\begin{figure}[htbp] 
   \centering
   \includegraphics[width=1.2in]{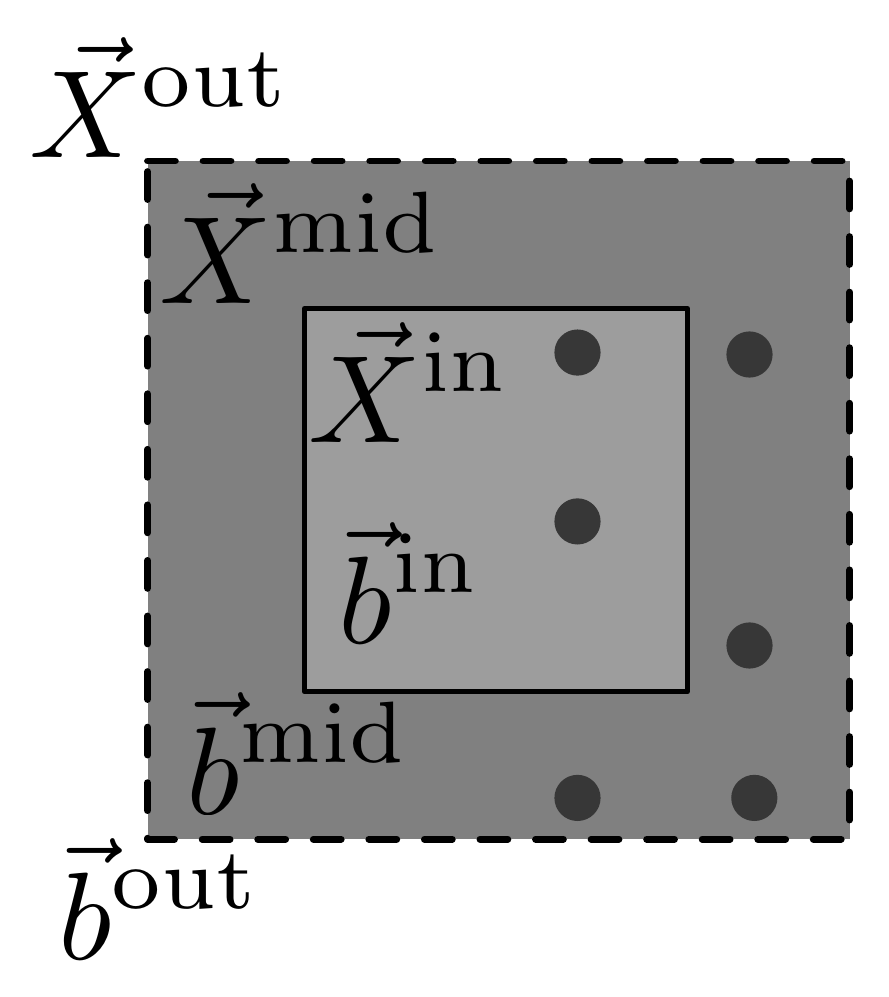} 
   \caption{The figure illustrates the definitions of random variables corresponding to bit-nodes and output (codeword) nodes at the encoder. The values $\vec{Y}^{\mathrm{in}}, \vec{Y}^{\mathrm{mid}}, \vec{Y}^{\mathrm{out}}$ are the counterparts of $\vec{X}^{\mathrm{in}}, \vec{X}^{\mathrm{mid}}, \vec{X}^{\mathrm{out}}$ viewed through the channel. It is important to note that they are not based on circuit partitioning at the \textit{decoder}. Indeed, for deriving bounds for the encoder $\bm$, we assume no implementation constraint on the decoder, so it is not even necessary that the decoder is implemented within the Implementation Model of Section~\ref{sec:implementation}. }
   \label{fig:encodingpic}
\end{figure}

We again choose the Stencil parameters $a=\frac{1}{\sqrt{2}}\sqrt{\frac{\log\frac{1}{10\peblk}}{\log\frac{1}{2p_{ch}}}}\lambda$ and $\eta=\frac{1}{4}$. Focusing on the $i$-th encoder subcircuit, let $n_i$ denote the number of codeword symbols inside the $i$-th encoder subcircuit, and let $k_i^\mathrm{inside}$ denote the input nodes (that store uncoded information) inside the inner square of the subcircuit. Further, for the $i$-th subcircuit, (dropping subscript $i$ for simplicity) let the information stored in the input nodes inside the inner square be denoted by $\vec{b}^\mathrm{in}$, and in those outside the outer square be $\vec{b}^\mathrm{out}$. There are more input nodes in the ``annulus'' between the inner square and the outer square, denote them by $\vec{b}^\mathrm{mid}$ (see Fig.~\ref{fig:encodingpic}). Similarly, define codeword symbols $\vec{X}^{\mathrm{in}}, \vec{X}^{\mathrm{mid}}, \vec{X}^{\mathrm{out}}$ and the corresponding channel outputs $\vec{Y}^{\mathrm{in}}, \vec{Y}^{\mathrm{mid}}, \vec{Y}^{\mathrm{out}}$ (see Section~\ref{sec:chmodel}). 

Now, at the decoder, declare the values of $\vec{X}^\mathrm{out}$ for free. Further, assume that the decoder is not required to recover the values of $\vec{b}^\mathrm{mid}$, $\vec{b}^\mathrm{out}$. Thus the job of the decoder is to only recover $\vec{b}^\mathrm{in}$ (this relaxation on requirements from the decoder will only further reduce the error probability). For recovering $\vec{b}^\mathrm{in}$, it has the channel outputs $\m{Y}$, and the freely declared vector $\vec{X}^\mathrm{out}$. Using the erasure-channel argument used in decoding lower bounds (Theorem~\ref{thm:decoding}), we assume that $\m{Z}$, the outputs of an erasure channel, are available at the decoder as well (which, as far as this theorem is concerned, is free to run the optimal Maximum Likelihood decoding without the constraints of implementation imposed on the encoder). This will only reduce the error probability for the same number of encoding $\bm$. Further, observing that $\vec{X}^\mathrm{out}$ are available to the decoder, we are interested in minimizing the entropy $H(\vec{b}^\mathrm{in}|\vec{X}^\mathrm{out},\m{Y},\m{Z}) $, which is the uncertainty at the decoder in the information bits (that are still undeclared, namely the information bits in the $i$-th encoder subcircuit) given the information available at the decoder to decode these bits. Examining this uncertainty,
\begin{eqnarray}
\label{eq:uncertain}H(\vec{b}^\mathrm{in}|\vec{X}^\mathrm{out},\m{Y},\m{Z}) &\overset{(a)}=& H(\vec{b}^\mathrm{in}|\vec{X}^\mathrm{out},\m{Z})\\
&\overset{(b)}=&H(\vec{b}^\mathrm{in}|\vec{X}^\mathrm{out},\vec{Z}^\mathrm{in},\vec{Z}^\mathrm{mid} ),
\end{eqnarray}
where $(a)$ and $(b)$ follow from the Markov chains $\vec{b}^{\mathrm{in}}\to\{\vec{X}^\mathrm{out},\m{Z}  \}  \to\m{Y} $ and $\vec{b}^{\mathrm{in}}\to  \{\vec{X}^\mathrm{out},\vec{Z}^\mathrm{in},\vec{Z}^\mathrm{mid}\}\to  \vec{Z}^\mathrm{out}$ respectively. 

Similarly, 
\begin{eqnarray}
\label{eq:uncertain2}
\nonumber && H(\vec{b}^\mathrm{in}|\vec{X}^\mathrm{out},\m{Y}=\m{y},\m{Z}=\m{z}) \\
\nonumber &=& H(\vec{b}^\mathrm{in}|\vec{X}^\mathrm{out},\m{Z}=\m{z})\\
&=&H(\vec{b}^\mathrm{in}|\vec{X}^\mathrm{out},\vec{Z}^\mathrm{in}=\vec{z}^\mathrm{in},\vec{Z}^\mathrm{mid}=\vec{z}^\mathrm{mid} ).
\end{eqnarray}
That is, the equality~\eqref{eq:uncertain} also holds for specific values of the random variables $\m{Y}$ and $\m{Z}$.


Our next step, which is key to this proof, is a simple equality. Consider the event that all of the symbols $\{\vec{Z}^\mathrm{in},\vec{Z}^\mathrm{mid}\}$ are erased, denoted by $\{\vec{Z}^\mathrm{in},\vec{Z}^\mathrm{mid}\}=E$. Then,
\begin{equation}
\label{eq:key}H(\vec{b}^\mathrm{in}|\vec{X}^\mathrm{out},\{\vec{Z}^\mathrm{in},\vec{Z}^\mathrm{mid}\}=E ) = H(\vec{b}^\mathrm{in}|\vec{X}^\mathrm{out}).
\end{equation}
This is because the event $\{\vec{Z}^\mathrm{in},\vec{Z}^\mathrm{mid}\}=E$ does not alter the joint distribution of $\vec{b}^\mathrm{in},\vec{X}^\mathrm{out}$ \textit{even when encoding is a flexible-message-length computation}. The encoder has no knowledge of this erasure-event\footnote{In absence of feedback from the receiver, the encoder only knows the channel statistics, not the realization. While feedback from the receiver to the transmitter is absent here, in presence of noiseless feedback, our bound on encoding $\overline\bm$\;could be beaten. But the question is more interesting and relevant with realistic models of noisy feedback, where benefits are severely curtailed (see, e.g.~\cite{LiFeedback}). Further, it is also important to note that for flexible-message-length implementations, the key equality~\eqref{eq:key} holds only when we are investigating circuits at the encoder. At the decoder, the knowledge that all inputs in the subcircuit are erased can be used by a subcircuit to ask for more information from the rest of the decoding circuit. At this point, it is unclear to us if this means that flexible-message-length decoding can beat our bound in Theorem~\ref{thm:decoding}.}, and thus cannot alter the joint distribution in response to the event. 
Further, under this erasure-event, because $\{\vec{Z}^\mathrm{in},\vec{Z}^\mathrm{mid}\}$ are completely erased, they provide no help in decoding $\vec{b}^\mathrm{in}$.

Thus, if $H(\vec{b}^\mathrm{in}|\vec{X}^\mathrm{out})\geq \frac{2k_i^\mathrm{inside}}{3}$ (as in~\eqref{eq:fano}), then the conditional probability of error in recovering these bits, $ \Pr(\vec{b}^\mathrm{in}\neq \vec{\widehat{b}}^\mathrm{in}|\{\vec{Z}^\mathrm{in},\vec{Z}^\mathrm{mid}\}=E)$, is at least $\frac{1}{9}$ (from Lemma~\ref{lem:fanoex}), and thus the (unconditional) block-error probability is lower bounded by
\begin{eqnarray}
\nonumber&&\peblk \geq \Pr(\vec{b}^\mathrm{in}\neq \vec{\widehat{b}}^\mathrm{in})
\\\nonumber&\geq& \Pr(\{\vec{Z}^\mathrm{in},\vec{Z}^\mathrm{mid}\}=E) \Pr(\vec{b}^\mathrm{in}\neq \vec{\widehat{b}}^\mathrm{in}|\{\vec{Z}^\mathrm{in},\vec{Z}^\mathrm{mid}\}=E)
\\&\geq& \frac{(2p_{ch})^{n_i}}{9},
\end{eqnarray}
 which leads to a contradiction (following the exact sequence of steps in~\eqref{eq:contra} from proof of Theorem~\ref{thm:decoding}).

Thus $H(\vec{b}^\mathrm{in}|\vec{X}^\mathrm{out})< \frac{2k_i^\mathrm{inside}}{3}$ for all $i$. This means that
\begin{eqnarray}
I(\vec{b}^\mathrm{in};\vec{X}^\mathrm{out}) = H(\vec{b}^\mathrm{in})  -H(\vec{b}^\mathrm{in}|\vec{X}^\mathrm{out}) \nonumber\\
> k_i^\mathrm{inside}- 2k_i^\mathrm{inside}/3 = k_i^\mathrm{inside}/3.
\end{eqnarray} 
Thus, at least $k_i^\mathrm{inside}/3$ bits of information are communicated from inside the inner square to outside the outer square for each subcircuit $i$ at the encoder. From Lemma~\ref{lem:fixedschedule}, the required $\bm$\;(average or deterministic) for the computation is at least $\eta a\frac{k_i^\mathrm{inside}}{3} = \frac{1}{12}k_i^\mathrm{inside}a$ (since $\eta=\frac{1}{4}$) for each subcircuit $i$ during encoding, and thus the total average $\bm$\;for encoding circuitry is at least $\frac{1}{12}k^{\mathrm{inside}}a=\frac{1}{48}ka$, yielding the lemma. 
\end{IEEEproof}
We emphasize that while our lower bounds for fixed and flexible-message-length encoding are the same, this does not imply that flexible-message-length cannot reduce the required energy consumption because our bounds could be loose. As we discuss in Section~\ref{sec:conclusions}, this necessitates a comparison with upper bounds, which is a work in progress.

\subsection{Lower bounds on \textbf{total} energy consumption}
\label{sec:total}
This section uses the bounds on \bm\;derived above to yield bounds on total (transmit and information-friction) energy consumed in communications. Strictly speaking, our bounds are for total energy-per-bit. However these bounds can be translated to total power consumption simply by dividing both transmission and circuit energy by the available time (under the assumption that encoding/decoding can take only as much time as transmission in order to not have buffer-overflows). The results in this section can be viewed as those that account for frictional losses in both the communication channel and the transmitter and receiver circuitry. However, our emphasis is on observing  qualitative differences between bounds on total energy and the traditional understanding on transmit energy. Thus we fix the distance (and hence also the path-loss) between the transmitter and the receiver, focusing on the contribution of circuit energy bounds to the total energy.

\begin{corollary}[Unavoidable limits on total energy-per-bit]
For communication over a channel with Channel Model ($\zeta,\sigma_z^2$) with the encoder and the decoder implemented in Implementation Model ($\lambda,\mu$) with fixed-message-length computing, the total energy per bit for communication at error probability $\peblk$ is lower bounded as:
 \begin{equation}
  \frac{E_{total}}{k}\geq \Omega\left(\sqrt[3]{\log\frac{1}{\peblk}}\right).
 \end{equation}
\end{corollary}
\begin{IEEEproof}
The lower bound  considers only the energy at the transmitting end: the transmit and the encoding energy, ignoring the decoding energy. This makes no difference to the order-sense result since the bounds in Theorem~\ref{thm:decoding} and Theorem~\ref{thm:encoding} are the same. 

Because the channel is used $W$ times per second, the per-bit transmit energy used is $\frac{nP_T}{W}$. The total (transmit + encoding) energy-per-bit under condition~\eqref{eq:condition} can therefore be lower bounded as (using Theorem~\ref{thm:encoding}, and denoting total transmit energy by $E_{Tx}$, and encoding energy by $E_{enc}$):
\begin{eqnarray*}
\frac{E_{total}}{k}&> & \frac{E_{Tx}+E_{enc}}{k}\\
&\geq & \frac{1}{k}\frac{n P_T}{W}+  \frac{1}{k}\frac{\mu k}{48\sqrt{2}}\sqrt{\frac{\log\frac{1}{10\peblk}}{\log\frac{1}{2p_{ch}}}}\\
&=& \frac{P_T}{RW}  + \frac{\mu}{48\sqrt{2}}\sqrt{\frac{\log\frac{1}{10\peblk}}{\log\frac{1}{2p_{ch}}}}.
\end{eqnarray*}
 In our hard-decision channel model, as $P_T$ increases, the term $\log{\frac{1}{2 p_{ch}}}$ scales proportionally to the received power  $\zeta P_T$ (see, e.g.~\cite{ISIT12Paper}). Thus
 \begin{eqnarray*}
 \frac{E_{total}}{k}&\geq &\frac{P_T}{RW}+ \frac{\beta}{48\sqrt{2}}\sqrt{\frac{\log\frac{1}{10\peblk}}{P_T}},
 \end{eqnarray*}
 for some $\beta>0$. By simple differentiation, the choice of $P_T$ that minimizes the RHS is $P_T^* = \Theta\left(\sqrt[3]{\log\frac{1}{\peblk}}\right)$. Substituting,
 \begin{equation}\label{eq:lower}
  \frac{E_{total}}{k}\geq \Omega\left(\sqrt[3]{\log\frac{1}{\peblk}}\right).
 \end{equation}
 If~\eqref{eq:condition} is not satisfied, then $P_T=\Omega\left(\log\frac{1}{\peblk}\right)$ (see \textit{Remark} after the statement of Theorem~\ref{thm:decoding}), which is larger than the behavior in~\eqref{eq:lower}. 
 \end{IEEEproof}
\textit{Remark}: While these bounds hold for any fixed communication distance in the limit of $\peblk\to 0$, it is important to note that for practically interesting values of $\peblk$ (typically between $10^{-3}$ and $10^{-20}$), empirical evidence~\cite{Ganesan2011,HowardSchlegel,Globecom12Paper} suggests that relative to transmit power, circuit power is relevant only at short distances (less than a few kilometers). At longer distances, the energy consumed in circuits at high $\peblk$ can be neglected in total power optimization because the transmit power is dominant. However, there can be situations where decoding power is still important because the receiver can be more energy constrained than the transmitter (e.g.~in the downlink of a cellular system). 

\subsection{What happens as the code-rate approaches the channel capacity?}

In practical situations, transmit power can be constrained by regulating authorities (e.g. the FCC) or the limit of the power amplifier at the transmitter circuitry. In such situations, it is not possible to increase transmit power to reduce the required encoding and decoding power. While our past work has shown that energy can be expended in other components (e.g. the equalizer or the beamformer) to effectively increase the SNR at the decoder~\cite{eusipcopaper}, thereby providing analogous  tradeoffs between transmit and circuit power as above, there likely are saturation-effects to such approaches as well (e.g. the thermal noise limit or interference due to ambient transmissions that are unaccounted for). 

What happens when the code rate is maintained near channel capacity (or, by keeping transmit power near Shannon limit for a fixed rate, the channel capacity is maintained near the code rate) even in the asymptotic limit of $\peblk\to 0$? Is the energy-cost higher than the case when we relax the constraint of operating close to capacity? Our earlier work shows this is the case~\cite{JSAC11Paper} for energy consumed in \textit{computational nodes} in the VLSI model (but does not show it for wiring energy, or the information-frictional energy for movement of information). Is this the case for information-frictional energy as well? The theorem below proves that this is indeed the case, and in fact, the information-frictional energy consumption is significantly higher (in order sense) than the energy consumed in computational nodes. The key observation used in the derivation of the following result is that small enough $\bm$\;in computation can lead to multiple sub-circuits having local decoding errors due to independent channel events. Because error in any one subcircuit leads to a block-error, and the error-events used to lower bound the error-probability of different subcircuits are independent, a stronger lower bound can be derived that captures a stronger dependance on $n$. 

\begin{theorem}
\label{thm:decodingn}
For an error correcting code transmitted over a channel with Channel Model ($\zeta,\sigma_z^2$) and decoded in a decoder circuit $\mathrm{DecCkt}$ implemented in Implementation Model ($\lambda,\mu$) with fixed-message-length implementation and block-error probability $\peblk$, the decoder \bm\;are lower bounded as:
\begin{equation}
\bm \geq \frac{k}{192}\sqrt{\frac{ \log n}{\log\frac{1}{2p_{ch}}}}\lambda,
\end{equation}
as long as 
\begin{equation}
\label{eq:conditionn}
\log n > 100\log\frac{1}{2p_{ch}}
\end{equation}
\end{theorem}
\begin{IEEEproof}
See Appendix~\ref{app:capacity}. 
\end{IEEEproof}
\textit{Remark}: The theorem shows that (under condition~\eqref{eq:conditionn}) as $n\to\infty$, the required $\bm$\;per-bit, \textit{i.e.} $\frac{\bm}{k}$ diverge to infinity as $\sqrt{\log n}$ for fixed transmit power. It is well known (e.g.~\cite[Exercise 5.23]{Gallager}\cite{verdudispersion}) that close to capacity, as $\peblk$ is made small for a fixed rate, $n\gtrsim \Omega\left(\frac{\log\frac{1}{\peblk}}{K(C-R)^2}\right)$ for some constant $K$ (that depends on $p_{ch}$). That is, the ``speed'' of increase of block length (and hence also of $\bm$\;per-bit) as $\peblk\to 0$ blows up as the code-rate approaches capacity.

Further, note that condition~\eqref{eq:conditionn} is satisfied in the asymptotic limit $\peblk\to 0$ for fixed-rate communication problems where communication is close to capacity. This is because in such situations, the transmit power needs to be maintained close to the Shannon limit (a constant at fixed rate), and thus $\log\frac{1}{p_{ch}}$ is bounded even as $\peblk\to 0$.

\section{Justification for, and the limitations of, the information-friction model}
\label{sec:examples}

\subsection{Practical examples where information-friction model applies}

\begin{figure}[htbp] 
   \centering
   \includegraphics[width=3.6in]{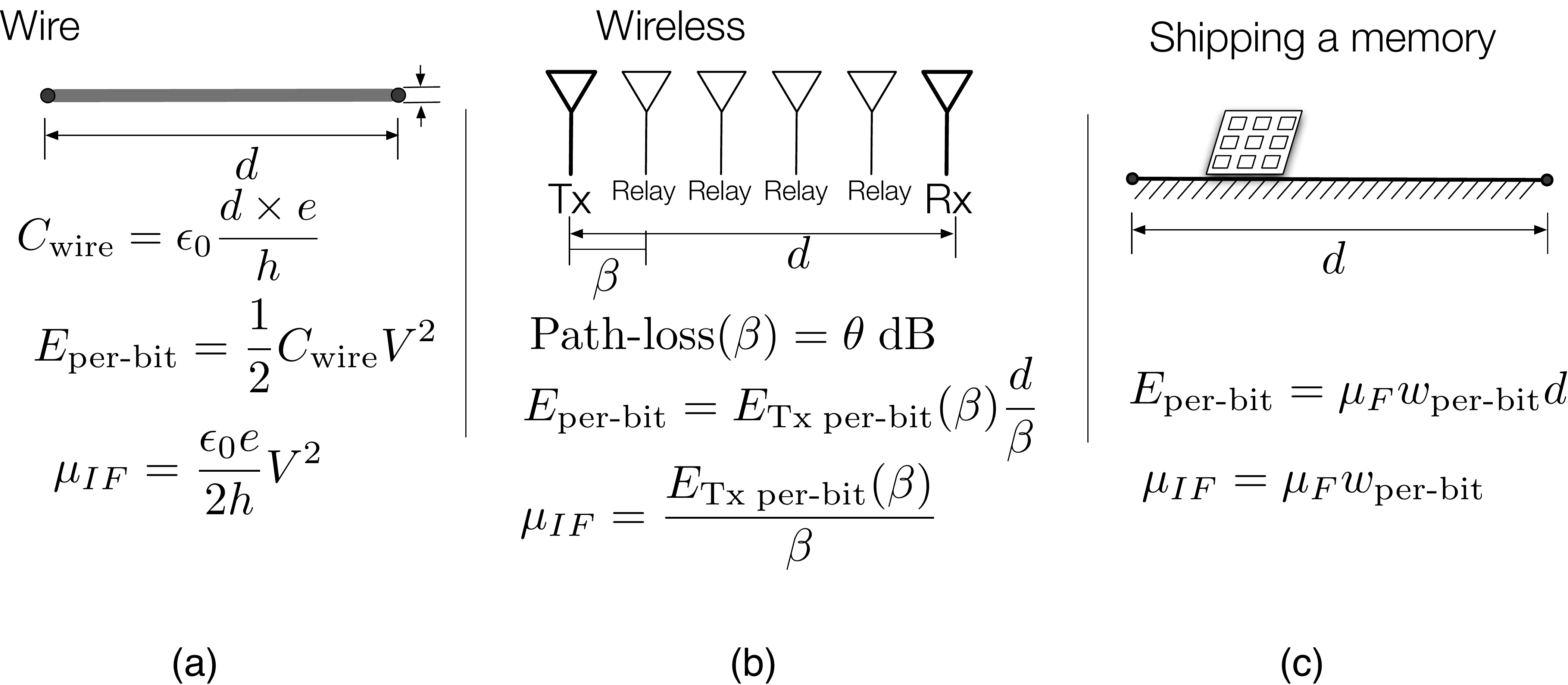} 
   \caption{Example ways of communication where the Information-Friction Model is a good approximation for energy consumption. The coefficient of information-friction, $\mu_{IF}$, is also calculated. For clarity, we use $\mu_{IF}$ for coefficient of information-friction here to distinguish it from $\mu_F$, the Newtonian coefficient of friction. In (a), $C_{\mathrm{wire}}$ is the capacitance of an on-chip wire of length $d$, width $e$, and distance $h$ from the computational substrate. $\epsilon_0$ is the permittivity of air. In (b), $\beta$ is the minimum possible distance between two repeaters, and $E_{\mathrm{Tx\;per}\text{-}\mathrm{bit}}(\beta)$ is the required energy per-bit to communicate to distance $\beta$. In (c), $w_{\mathrm{per}\text{-}\mathrm{bit}}$ is the weight of a memory cell that stores a bit.}
   \label{fig:examples}
\end{figure}

The following modes of communicating (via binary symbols\footnote{The binary-symbol-assumption is made for simplicity. The results can easily be extended for any fixed ``constellation size.''}) in a computational system motivate our definition of information-friction:

\textbf{Metal wires}: A metal wire can be modeled as a capacitance between the wire and the substrate (often referred to as the Elmore lumped model~\cite{JanBook}) that needs to be charged and discharged through the wire resistance (see Fig.~\ref{fig:examples} (a)). The amount of energy expended by a single charge/discharge operation (e.g., to communicate one bit of information) is $\frac{1}{2} C_{\mathrm{wire}}V^2$~\cite{JanBook}, which increases linearly with the wire-capacitance (here $V$ is the voltage across the wire). Further, the capacitance $C_{\mathrm{wire}}=\epsilon\frac{d\times e}{h}$ itself increases linearly with the wire-length $d$ (here $e$ is the wire-width, and $h$ is the distance of the wire from the substrate). Thus each charge/discharge requires energy that scales linearly with the wire-length. The resulting ``coefficient of friction'' is shown in Fig.~\ref{fig:examples} (a).

\textbf{Wireless links}: While wireless communication from a single transmitting node to a single receiving node suffers from worse-than-linear losses (path-loss is often $\frac{1}{d^\psi}$ for some $\psi>2$), with repeaters placed at uniform distances, the energy requirement can be brought down to approximately linear with distance. However, this is only possible when the signal-to-noise ratio is large, which is when the errors are so improbable that their propagation along the relays can be neglected.  When these errors cannot be neglected, or when the repeaters are difficult to place, the information-friction model only provides a loose bound on the total energy. 

\textbf{Transporting matter}: One way of communicating is via writing the message on a memory, and sending the memory from the sender to the receiver~\cite{RoseShip}. Not only is this form of communication widely used today (e.g.~use of USB sticks and CDs to transport information), it has been envisaged as a method of communication in a billiard-balls computer by the physicists~\cite{ToffoliBilliard}, and has also been thought of as an efficient way of communicating across interstellar space~\cite{RoseShip}. Any friction (e.g.~friction between the medium of transport, such as the billiard ball, and the surface) that the transported object faces gets translated into information-friction via the weight of one bit of storage unit. It appears that pneumatic computers (that store and communicate between logic elements using fluid movements, e.g.~\cite{pneumatic}) would encounter similar frictional losses. For fluid traveling through smooth pipes, there is still a loss in pressure which is linear per-unit length (see~\cite{FrictionFactor}). For communicating computational messages reliably, this pressure loss will again necessitate use of repeaters, thereby leading to information-frictional losses just as those for wireless links.

\subsection{Limitations of the information-friction model}

At extremely low speeds of computation, it may be possible to reduce the coefficient of information-friction, consistent with results in thermodynamics of computation~\cite{LandauerComputation,BennettReversibility,BennettNotes}, by communicating using timing of the signal (such as in pulse-position modulation, or through silence~\cite{OrlitskySilence}). In such situations, with a single change in the message on a circuit-link, a large number of bits can be communicated (depending on how slow the computation can be). However, such techniques are hard to implement because they require sophisticated synchronization between circuit components in order to exploit communication via timing. Often this synchronization is performed by explicitly sending a clock-signal~\cite{JanBook}, and the communication of clock-signal itself can consume significant amount of energy. Thus it is unclear if communication using timing is a practical way to reduce the coefficient of information-friction significantly. 

While in most situations, information-friction bounds are valid (if loose) lower bounds on energy-consumption, we note that there could be situations where these bounds are beaten. One such situation is when a computation uses wireless broadcast for transmission on computation links. It is plausible, for instance, that when multicasting to multiple nodes simultaneously, the required energy can increase slower than linearly with the cumulative distance of communication. There is literature that uses broadcast as a way to reduce communication requirements in the sense of traditional (Andrew Yao's) communication complexity of distributed sorting~\cite{KleinrockBroadcast}. A deeper exploration is needed to understand if energy requirements can also be lowered for such computations via broadcast to beat the information-friction limits. 

Finally, we note that information-frictional energy is not always the dominant sink of energy in computational systems. While asymptotically, our theoretical results here and empirical observations in~\cite{Allerton12Paper,KarthikJournalPaper} strongly suggest that information-friction is the dominant sink, in practical systems, energy consumed in computational nodes or memory-access could be significant, and could even dominate in non-asymptotic scenarios. Improved modeling of energy consumed in nodes and memory-access could enhance the understanding in such scenarios.

\section{Discussions and conclusions}
\label{sec:conclusions}

The information-friction model proposed here can be viewed as a broadening and a simplification of the VLSI model introduced by Thompson and others. The model enjoys several advantages over the VLSI model. In particular, it can capture energy requirements in wired as well as wireless computational systems, and has a closer connection to energy consumption (as noted in the introduction). Within information-theoretic literature, our metric of \bm\;for computational costs has been used earlier as a metric for transport capacity of wireless networks~\cite{GuptaKumar}. Within physics, it has a potential connection with thermodynamics. Most of the classical analysis focuses on energy of single operations (e.g.~\cite{LandauerCompute,LandauerComm}), and even this analysis becomes difficult when the computation needs to be performed in non-infinite time\footnote{Finite-time analyses need to tackle non-equilibrium thermodynamics, which has proven to be quite hard (e.g.~\cite{Gawedzki}).}, in part because friction can no longer be ignored\footnote{Friction can be ignored in infinite-time analysis because changes can be made at speeds approaching zero, keeping the system in equilibrium at all times, lowering frictional losses to as low as desired.}. Recent works~\cite{ManojPaper1,ManojPaper2,Gawedzki} have shown promise towards addressing finite-time single-operation computing, but even once this is understood, it will still remain to extend the analysis to multi-operation computation. While our techniques here are guided strongly by current implementations, they could complement the single-operation-based analysis in statistical physics, offering suggestions regarding what form the fundamental limits should look like. 

Nevertheless, we do believe that an even broader approach is needed to understand how physically-fundamental our limits on energy are. The approach proposed here is not in the spirit of Landauer's, where the goal is to relax all constraints (timing of computation, frictional energy, medium of implementation, etc.) in obtaining fundamental limits. Instead, this approach is closer to Shannon's engineering approach: just as Shannon modeled the communication channel and derived fundamental limits that hold for all possible communication strategies \textit{for the chosen channel model}, here we model the communication channel \textit{and the implementation}, and derive limits that hold for all possible communication strategies and implementation architectures and algorithms \textit{for the chosen implementation model}. The key assumptions lie in modeling of implementation, and a good first step towards deeper understanding can be to relax or modify these assumptions\footnote{As Norbert Wiener noted on choice of assumptions, ``What most experimenters take for granted before they begin their experiments is infinitely more interesting than any results to which their experiments lead.''}.

Are these limits useful in guiding code-design? Our complementary work with experimentalists~\cite{Ganesan2011,Globecom12Paper} that provides upper bounds on energy has shown that the code-choice needs to adapt to distance of communication: at shorter distances, simpler coding techniques (that require smaller wire-length per-bit) are more total-energy-efficient than capacity-approaching codes. As distances of communication increase, approaching capacity becomes increasingly efficient. In this paper, for reasons of clarity, we have fixed the communication distance (see Section~\ref{sec:chmodel}). Even for purely intellectual reasons, it is important to explore these upper bounds further and obtain an order-sense asymptotic understanding (along the lines of~\cite{Allerton12Paper}) on how tight the lower bounds are, and if the suggestions we draw via comparison of lower bounds (e.g.~using bounded transmit power as $\peblk\to 0$ fundamentally requires larger total power) in this paper actually hold. 

One also needs to understand the implications in multi-user situations, especially in interference-limited situations where the advantage of increasing transmit power indefinitely can be limited by saturation of SINR, as explored in~\cite{JSAC11Paper}. Intuitively in such situations~\cite{JSAC11Paper}, as the density of transmitting devices increases, it becomes increasingly important to save transmit power (that can cause interference) even at the cost of increased encoding and decoding energy. It might be the case that energy-efficient radios need to be ``cognitive'' in detecting nearby transmitter and receiver density, and choosing the optimal energy-efficient strategy in response. 
 
Finally, an important question remains to be understood in the total energy of point-to-point communication: how much can feedback help? Perfect (noiseless, infinite-precision) feedback can help in reducing complexity significantly~\cite{Kailath66}. However, perfect feedback is impossible to obtain in practice, and more reliable feedback also requires an increased energy cost (just as more reliable forward transmission does). One will therefore need to examine the issue in presence of noisy feedback, of which the understanding is far from mature, especially from a fundamental-limits perspective (e.g.~\cite{kimNoisyFeedback,KimFeedback2,burnashevZeroRate1,burnashevZeroRate2,LiFeedback}). More broadly, we also need to allow noise in the computation process itself (some of our recent work, e.g.~\cite{ISIT14Paper,YaoqingAllerton14}, focuses on this issue), a line of work started by von Neumann~\cite{vonNeumann} that still lacks a strong connection with energy consumption.

\section*{Acknowledgments}

We acknowledge the generous support of NSF grants NSF-ECCS-1343324, NSF CAREER (NSF-CCF-1350314), and a startup grant from Carnegie Mellon University. We also thank Manoj Gopalakrishnan and Sanjoy Mitter for helpful discussions, the reviewers for their detailed reading and helpful suggestions, and Majid Mahzoon and Yaoqing Yang for carefully reading and commenting on the final version.

\appendices

\section{Increase in decoding energy on approaching capacity}
\label{app:capacity}
This Appendix provides the proof of Theorem~\ref{thm:decodingn}.
\begin{IEEEproof}
Choose the Stencil parameter $a=\frac{1}{2}\sqrt{\frac{ \log n}{\log\frac{1}{2p_{ch}}}}\lambda$ for some $\xi<1$. Then, under condition~\eqref{eq:conditionn} (which guarantees that $\frac{a^2}{\lambda^2}= \frac{\log n}{4\log\frac{1}{2p_{ch}}}>\frac{100\log\frac{1}{2p_{ch}}}{4\log\frac{1}{2p_{ch}}}=25$, satisfying the condition of Lemma~\ref{lem:choosinga}), by Lemma~\ref{lem:choosinga}, $n_i\leq \frac{2a^2}{\lambda^2}= \frac{\log n}{2\log\frac{1}{2p_{ch}}}$.

The rest of the proof uses ideas from the work of Blake and Kschischang~\cite{BlakeKschischang} to bound block-error probability under independent subcircuit error events, and is via contradiction. Choose $\eta=\frac{1}{4}$, and suppose $\bm < \frac{k}{192}\sqrt{\frac{ \log n}{\log\frac{1}{2p_{ch}}}}\lambda = \frac{1}{24}k\eta a\leq \frac{1}{24}k^{\mathrm{inside}} a$ (for appropriately chosen Stencil origin). Under this assumption, we first claim (and prove via contradiction) that for at least $\frac{k^{\mathrm{inside}}}{2}$ bit-nodes, the subcircuits that they lie in have $\bm_i\leq \frac{k_i^{\mathrm{inside}}}{12} a$. Suppose our claim is not correct. Then for at least $\frac{k^{\mathrm{inside}}}{2}$ bits, the subcircuits they lie in have $\bm_i> \frac{k_i^{\mathrm{inside}}}{12} a$, which would mean that the total number of \bm\;is larger than $\frac{k^{\mathrm{inside}}}{24} a$, leading to a contradiction. Thus at least $\frac{k^{\mathrm{inside}}}{2}$ bit-nodes lie in subcircuits with $\bm_i\leq \frac{k_i^{\mathrm{inside}}}{12} a$. With $\eta=\frac{1}{4}$, this means that at most $\frac{k_i^{\mathrm{inside}}}{3}$ bits of information is available to decode these $k_i^{\mathrm{inside}}$ bits in the event of erasure of all the channel outputs inside the outer square of the $i$-th subcircuit, leading to a lower bound of $\frac{1}{9}$ on error probability conditioned on this erasure event.  

Now notice that at the decoder, these erasure events are independent across different circuits. Further, the information inside \textit{every} subcircuit needs to be recovered in order to recover the entire block. This yields the following stronger lower bound on the block-error probability. 
\begin{equation}
\label{eq:newlb}
\peblk \geq 1- \prod_{i:\bm_i\leq \frac{k_i^{\mathrm{inside}}}{12} a}\left(1-\frac{(2 p_{ch})^{n_i}}{9}\right).
\end{equation}
where the set $Err:=\{i:\bm_i\leq \frac{k_i^{\mathrm{inside}}}{12} a\}$ is the set of subscript-indices such that each such subcircuit has error probability in recovering its information bits lower bounded by $\frac{(2 p_{ch})^{n_i}}{9}$. Because at least $\frac{k^{\mathrm{inside}}}{2}$ number of bits lie in subcircuits with $ \bm_i\leq \frac{k_i^{\mathrm{inside}}}{12} a$, and from Lemma~\ref{lem:choosinga}, $k_i^{\mathrm{inside}}\leq \frac{2a^2}{\lambda^2}=\frac{ \log n}{2\log\frac{1}{2p_{ch}}}$ for any subcircuit, it has to be the case that 
\begin{equation}
|Err|\geq \frac{\frac{k^{\mathrm{inside}}}{2}}{{\frac{ \log n}{2\log\frac{1}{2p_{ch}}}}}\overset{(a)}\geq \frac{\frac{k}{4}}{{\frac{ \log n}{\log\frac{1}{2p_{ch}}}}}= \frac{nR \log\frac{1}{2 p_{ch}}}{4\log n},
\end{equation}
where $(a)$ uses the fact that $k^{\mathrm{inside}}\geq (1-2\eta)^2k=\frac{k}{4}$ (since $\eta=\frac{1}{4}$). Thus,
\begin{equation}
\label{eq:newlb2}
\peblk \geq 1- \left(1-\frac{(2 p_{ch})^{\bar{n}}}{9}\right)^\frac{nR \log\frac{1}{2 p_{ch}}}{4\log n},
\end{equation}
where $\bar{n}:=\frac{ \log n}{2\log\frac{1}{2p_{ch}}}$ is also an upper bound on $n_i$ for each $i$. Examining the second term in the RHS of~\eqref{eq:newlb2} by taking its $\log$,
\begin{eqnarray*}
&&\log\left(1-\frac{(2 p_{ch})^{\bar{n}}}{9}\right)^{\frac{nR \log\frac{1}{2 p_{ch}}}{4\log n}}\\
&=&{\frac{nR \log\frac{1}{2 p_{ch}}}{4\log n}}\log\left(1-\frac{(2 p_{ch})^{\bar{n}}}{9}\right)\\
&= & \frac{nR \log\frac{1}{2 p_{ch}}}{4\log n}\log\left(1-\frac{(2 p_{ch})^{\frac{ \log n}{2\log\frac{1}{2p_{ch}}}}}{9}\right)\\
&= & \frac{nR \log\frac{1}{2 p_{ch}}}{4\log n}\log\left(1-\frac{1}{9n^\frac{1}{2}}\right)\\
&\approx & \frac{nR \log\frac{1}{2 p_{ch}}}{4\log n}\left(-\frac{1}{9\sqrt{n}}\right)\\
&\overset{n\to \infty}\to& -\infty.
\end{eqnarray*}
Thus,  the second term in the RHS of~\eqref{eq:newlb2} goes to 0, and $\peblk\to 1$ as $n\to\infty$, leading to a contradiction.
\end{IEEEproof}

\bibliographystyle{IEEEtran}
\bibliography{IEEEabrv,MyMainBibliography,MyMainBibliography2}


\end{document}